%% file: main.tex
\documentclass[10pt, conference, letterpaper]{IEEEtran}
\IEEEoverridecommandlockouts
\usepackage{cite}
\usepackage{amsmath,amssymb,amsfonts}
\usepackage{graphicx}
\usepackage{textcomp}
\usepackage{xcolor}
\usepackage{pifont}
\def\BibTeX{{\rm B\kern-.05em{\sc i\kern-.025em b}\kern-.08em
    T\kern-.1667em\lower.7ex\hbox{E}\kern-.125emX}}
\usepackage{tikz}
\usepackage{amsmath}
\usepackage{array}
\usepackage{booktabs}
\usepackage{multirow}
\usepackage{makecell}
\usepackage[noend,ruled,boxed,commentsnumbered,linesnumbered]{algorithm2e} 

\usepackage{hyperref}

\begin{document}

\title{NebulaFL: Effective Asynchronous Federated Learning for JointCloud Computing
}
\author{
\IEEEauthorblockN{Fei Gao$^1$, Ming Hu$^2$*, Zhiyu Xie$^1$, Peichang Shi$^1$*, Xiaofei Xie$^2$, Guodong Yi$^3$, Huaimin Wang$^{1}$}
\IEEEauthorblockA{\textit{$^1$National Key Lab. of Parallel and Distributed Processing, National University of Defense Technology, Changsha, China} \\
\textit{$^2$School of Computing and Information Systems, Singapore Management University, Singapore}\\
 \textit{$^3$Xiangjiang Lab,Changsha, China}
}
\thanks{* Ming Hu (\url{hu.ming.work@gamil.com}) and Peichang Shi (\url{pcshi@nudt.edu.cn}) are corresponding authors}
}

\maketitle

\input{abstract}

%第1章  介绍
\input{intro}
\input{related_work}

%第3章 系统简介
\input{approach}

% %第4章 设计
% \section{Design}
% XXXX

% %第5章 收敛性分析
% \section{CONVERGENCE ANALYSIS}
% XXXX

%第6章 实验评估
\input{experiment}

%第8章 conclusion
\input{conclusion}

% \section*{Acknowledgments}
% This work was supported by National Key R\&D Program of China(No.2022ZD0115302), in part by the National Natural Science Foundation of China (No.62202479, No.61772030), the Open Fund of PDL(No. 2023-KJWPDL-06), the Major Program of Xiangjiang Laboratory (No.22XJ01004) and the Major Project of Technology Innovation of Hunan Province (No.2021SK1060-1).

%\begin{thebibliography}{1}

\bibliographystyle{IEEEtran}
%\bibliographystyle{elsarticle-num}

% \clearpage
% Loading bibliography dablebase
\bibliography{refs}

\end{document}

%% file: abstract.tex
\begin{abstract}
With advancements in AI infrastructure and Trusted Execution Environment (TEE) technology, Federated Learning as a Service (FLaaS) through JointCloud Computing (JCC) is promising to break through the resource constraints caused by heterogeneous edge devices in the traditional Federated Learning (FL) paradigm.
Specifically, with the protection from TEE, data owners can achieve efficient model training with high-performance AI services in the cloud. By providing additional FL services, cloud service providers can achieve collaborative learning among data owners.
However, FLaaS still faces three challenges, i.e., i) low training performance caused by heterogeneous data among data owners, ii) high communication overhead among different clouds (i.e., data centers), and iii) lack of efficient resource scheduling strategies to balance training time and cost.
To address these challenges, this paper presents a novel asynchronous FL approach named NebulaFL for collaborative model training among multiple clouds.
To address data heterogeneity issues, NebulaFL adopts a version control-based asynchronous FL training scheme in each data center to balance training time among data owners.
To reduce communication overhead, NebulaFL adopts a decentralized model rotation mechanism to achieve effective knowledge sharing among data centers.
To balance training time and cost, NebulaFL integrates a reward-guided strategy for data owners selection and resource scheduling. 
The experimental results demonstrate that, compared to the state-of-the-art FL methods, NebulaFL can achieve up to 5.71\% accuracy improvement.
In addition, NebulaFL can reduce up to 50\% communication overhead and 61.94\% costs under a target accuracy.

% As a privacy-preserving machine learning paradigm, Federated Learning (FL) is promising to facilitate collaborative learning among data silos.
% However, the performance of traditional central cloud server-based FL paradigm is seriously limited by edge devices.
% With the improvement of the AI infrastructure and Trusted Execution Environment (TEE) technology, cloud-based FL as a Service (FLaaS) is promising in large-scale cross-domain distributed computing, i.e., Joint-Cloud Computing.

\end{abstract}


%% file: intro.tex
\section{Introduction}

% 介绍大背景
% 目前AI模型发展已经被应用到医药、对话、金融、教育等，随着AI模型的迅速发展，使用更多的数据和资源将有助于实现模型效果，然而数据隐私成为一个关键的问题。因此联邦学习作为一个新兴的技术可以实现满足隐私保护的情况下训练一个全局模型with 共享数据，然而传统的联邦学习存在设备资源限制与网络通信等问题，能耗难以支持。随着AI基础设施及TEE技术的发展，很多公司在保护隐私的i情况下在云中存储数据，有效的利用各数据中心的大量的数据与资源进行来联邦学习提供服务（FLaaS）有希望打破传统联邦学习的资源局限，同时，数据中心拥有大量的数据，有助于提升模型质量，实现训练质量与训练速度的双赢。

% As Artificial Intelligence (AI) strategies demonstrate powerful advantages in various tasks such as image classification, text generation, and decision control, their deployment spans across multiple industry applications, including transportation~\cite{han2022ads}, healthcare~\cite{panesar2019machine}, and system design~\cite{hu2021enumeration}. 
% Typically, high-quality AI models require vast amounts of data for training. 
% However, data privacy concerns and legal restrictions make data collection increasingly challenging.
% To address privacy concerns, 

Due to the advantages of privacy protection, Federated Learning (FL)~\cite{mcmahan2017communication, li2020federated, hu2024fedmut} has emerged as a solution for distributed privacy-preserving machine learning. In the conventional FL paradigm, a central cloud server coordinates with multiple local clients for collaborative model training, where the cloud server dispatches a global model to clients for local training and aggregates models uploaded from clients to update the global model.
% In this way, FL can achieve distributed model training without data sharing.
Although FL has drawn a blueprint to address the problem of data silos, its practical application still faces significant limitations due to constraints in hardware resources and the uncertain physical environments of edge devices.
Specifically, personal edge devices, such as smartphones and IoT devices, often have limited computational power, memory, and storage, making it difficult for them to handle complex model training tasks. Furthermore, various uncertain factors~\cite{hu2020quantitative}, such as network delays, human-in-the-loop variability, and device failures, contribute to uncontrollable training processes and system fragility.

With the improvement of AI infrastructure, cloud-based AI training is becoming increasingly popular due to its superior efficiency and more stable performance compared to edge devices. 
Moreover, advances in Trusted Execution Environment (TEE)~\cite{sabt2015trusted} technology have effectively addressed concerns about data privacy leakage, making cloud-based model training more secure.
Leveraging cloud services and TEEs, data owners can securely upload their data to the cloud and perform FL using cloud-based AI services.
This has given rise to Federated Learning as a Service (FLaaS), a promising new offering for cloud providers that supports privacy-preserving collaborative training among different data owners.
Specifically, each data owner maintains a TEE container in the cloud and uploads their raw data to this container. The local training process is conducted within the TEE container to prevent privacy leakage. Cloud service providers then offer model aggregation and distribution services, ensuring a secure and efficient FL training process.

However, FLaaS still faces three serious challenges, i.e., \ding{182} heterogeneous data, \ding{183} cross-cloud training, and \ding{184} resource scheduling, respectively.
First, similarly to the traditional FL paradigm, FLaaS encounters low training performance due to the heterogeneity of data in different TEE containers, which cannot be shared.
Specifically, the data among data owners are non-Independent and Identically Distributed (non-IID), leading to the ``client drifting'' problem~\cite{karimireddy2020scaffold} that significantly limits the inference performance of the trained global model. Additionally, variations in data volume among data owners result in different local model training times, causing a straggler problem where slower clients delay the overall training process. 
Therefore, data heterogeneity remains a serious challenge in FLaaS.
Second, FLaaS typically involves collaborative training across multiple data centers, known as JointCloud computing~\cite{wang2017jointcloud,shi2017corporation}. Data owners are often located in different countries or regions. Due to legal constraints and the need for efficient data storage and transmission, even a single cloud service provider often utilizes multiple data centers. In the JointCloud scenario, FL requires frequent aggregation of local models from all participants, making communication overhead across these data centers a critical factor affecting the performance of FLaaS.
Furthermore, unlike traditional FL, cloud services usually involve a wide variety of computing resources. Balancing training time and the cost of renting computing resources while ensuring model training accuracy has become another significant challenge for FLaaS.

Unfortunately, most of the existing FL methods still focus on traditional cloud-edge architectures.
To address non-IID issues, mainstream FL methods adopt various strategies, such as global variables~\cite{karimireddy2020scaffold,li2020federated}, knowledge distillation~\cite{lee2022preservation}, client clustering~\cite{lin2022fedcluster}, or multi-model collaborative training~\cite{hu2024fedcross,hu2024aggregation} to enhance training performance.
To address the straggler issue, existing works attempt to utilize an asynchronous~\cite{xie2019asynchronous,gitfl2023} or simi-aynchronous~\cite{wu2020safa,ma2021fedsa} training mechanism.
For example, GitFL~\cite{gitfl2023} adopts multiple branch models instead of a global model for local training and performs a specific weighted aggregation (i.e., model pulling and model pushing) with the aggregated model (i.e., master model) for the knowledge sharing among branch models.
By assigning a low-version branch model to a high-performance client, GitFL can balance the training times of each branch model.
Although these methods can effectively improve the performance of conventional FL, they cannot be directly adapted to JointCloud scenarios. Specifically, the requirement of aggregation in each FL round results in unbearable communication overhead among data centers. 
Furthermore, existing methods do not consider resource scheduling, which prevents the effective utilization of the various computing resources available in data centers. This limitation particularly affects the ability to balance training time and cost.
Therefore, {\it how to perform FL in JointCloud scenarios wisely to balance training time and cost while ensuring training accuracy has become an important challenge for FLaaS.}

To address the following problems, the goal of this paper is to design \ding{182} an asynchronous training strategy in each data center to achieve effective training among its data owners, \ding{183} a communication-efficient knowledge-sharing mechanism among data centers, and \ding{184} a resource scheduling strategy to adaptively allocate resources based on the duration of training and resource rent.
Inspired by GitFL~\cite{gitfl2023}, to address the problem of data heterogeneity, each data center can maintain multiple intermediate models.
It selects data owners in real-time for local training of each intermediate model based on the version of the model and the historical or predicted training time of data owners under specific resources.
To reduce communication overhead among data centers, intuitively, each data center can request the aggregated model for other data centers at specific time intervals and use a weighted aggregation strategy to update their intermediate models to achieve knowledge sharing rather than real-time broadcasting of their intermediate models.
To balance the training time and cost of resource rental, intuitively, the scheduling strategy can prefer to assign high-performance resources for the training of low-version models and assign low-cost resources for that of high-version models.

Based on the motivation above, this paper presents a novel asynchronous FL approach named NebulaFL to support efficient collaborative machine learning in JointCloud scenarios.
Specifically, NebulaFL supports the architecture of multiple data centers, and each data center includes multiple data owners.
In NebulaFL, each data center maintains a list of intermediate models (named planet models) for asynchronous FL training and a sharing model (named stellar model) that is requested for other data centers or aggregated by all its planet models used for knowledge sharing.
In NebulaFL, to achieve communication-efficient knowledge sharing among data centers, each data center requests an aggregated model from a specific data center to update its stellar model and applies weighted aggregation between the updated stellar model and all its planet models for knowledge sharing.
NebulaFL adopts a version-aware strategy to select a container of a specific data owner for local training of each planet model and schedules hardware resources according to the model version and the predicted training time of the data owner.
In summary, this paper has the following four main contributions: 
% planet
% stellar
\begin{itemize}
    \item We present a novel asynchronous FL framework named NebulaFL for JointCloud scenarios, which supports FL among multiple data centers.
    \item We propose a communication-efficient knowledge sharing mechanism for the training of planet models among different data centers.
    \item We present a reward-based data owner selection and resource scheduling strategy to balance the training time and resource cost.
    \item We conduct extensive experiments to evaluate the performance of NebulaFL.
\end{itemize}

%% file: related_work.tex
\section{Preliminary and Related Work}

\subsection{Preliminaries}
%联邦学习经典的架构是一个中心服务器架构，注意到本篇论文关注多中心联邦学习下新场景，每个数据中心进行一个局部联邦学习训练，共同聚合为一个全局模型。  参考目前最典型的fedavg的聚合方法 
{\bf Federated Learning.}
The classic FL systems typically employ a cloud-edge architecture~\cite{li2014scaling}, which consists of a centralized server for model aggregation and multiple edge nodes for local training. 
However, this paper focuses on a new scenario involving multi-center FL. 
Consider that the FL System has a set of $\mathcal{C} = \left \{ C_{1}, C_{2}, \dots, C_{N}  \right \}$ data centers. Each datacenter $C_{i}$ has $M_{i}$ data owners holding hetergeneous data, which is denoted by  $D_{i} =\left \{ D_{i,1}, D_{i,2}, \dots , D_{i,M_{i}}\right \} $. In this setup, each data center trains its own local model with heterogeneous data owners, and these local models are collaboratively aggregated into a global model. So far, traditional federated learning methods commonly aggregate local models using the FedAvg algorithm~\cite{fedavg2019} to produce the global model. The objective of a FL optimization problem is typically defined as follows:

\begin{equation}
% \footnotesize
\begin{split}
    \min_w F(w) &= \frac{1}{N} \sum_{i=1}^N f_i(w) \\
    \text{s.t., } f_i(w) &= \frac{1}{M_i} \sum_{j=1}^{M_i} l(w; d_{i,j}).
\end{split}
\end{equation}

Here, $l$ and $f_i$ represent the loss function of a data owner's sample and all samples of data center $C_{i}$, and $F$ represents the global model's loss function. We focus solely on the inference performance of the unified global model for FLaaS, excluding personalized and clustered FL scenarios.

%B.现有优化方案。  有同步异步两个角度
%异步 自适应设备，现有的  同步方法设计  however 无法匹配我们的场景FLaaS 
%异步  gitFL 无法直接适配云际 只涉及单云内部
\subsection{Related Work on FL Optimization}

%为了提高传统联邦学习的推理性能，很多架构以及选择优化方案已经被大量研究。从架构上来讲，主要分为同步训练与异步训练方式优化，从方案优化来讲，主要包括效率提升，客户端选择，训练成本，通信优化等方面

To enhance the inference performance of traditional FL, numerous works about architectures and workflow optimization have been extensively researched, including edge-cloud collaboration, data heterogeneity management, resource allocation and task scheduling, communication optimization~\cite{li2022fedrelay,liu2020privacy}. So far, to improve the performance of specific FL methods, especially in non-IID scenarios, current work can be broadly categorized into the following three categories.
Firstly, the data heterogeneity problem is often addressed using techniques such as model regularization and data clustering to mitigate the impact of differences in data distribution on performance\cite{li2023distributed,liao2023adaptive,rodio2023federated}. For example, the method\cite{li2023distributed} mixes model updates from previous rounds with the current round's updates to avoid model shifts. In\cite{liao2023adaptive} proposed FedHP to achieve fast convergence by jointly optimizing both the local updating frequency and network topology in DFL.
Secondly, to tackle the straggler problem, various acceleration techniques have been developed\cite{xu2021helios,jiang2022model,ma2021fedsa,gitfl2023}. In synchronous FL, For example, PruneFL\cite{jiang2022model} introduced a pruning-based method designed to accelerate local training for stragglers in a synchronous setup. For asynchronous FL, Ma et al.\cite{ma2021fedsa} proposed a semi-asynchronous approach with a model buffer for aggregation. In gitFL\cite{gitfl2023}, a branch model with an older version is more likely to be assigned to a faster and less frequently selected device for the next round of local training. However, the above methods are not well-suited for FLaaS scenarios. The network differences within and between data centers can significantly increase the waiting time due to delays. Lastly, communication bottleneck is also a key issue hindering the performance of FL\cite{chen2023synchronize,li2022fedrelay,liu2020privacy,cui2023user,liu2023communication,guan2023enabling}. Work\cite{chen2023synchronize} proposes a new approach called Adaptive Parameter Freezing by freezing stable parameters intermittently during periods of synchronization. In\cite{cui2023user}, a lightweight deep neural network that adopts the designed lightweight fire modules and has a side branch for communication.

%尽管已经提出了很多优化方法来提升到联邦学习性能，但是由于智算中心的资源与网络情况不同，大部分方法无法直接应用到FL的场景，无法充分保证性能与资源的有效利用，因为我们提出了NebulaFL，在多云间采用双层架构在多数据中心之间进行联邦学习，以提供强大的推理服务。
Although many optimization methods have been proposed to enhance FL performance, they are mostly tailored to specific edge-cloud scenarios, which cannot be directly applied to JointCloud scenarios for due to varying resource availability and network conditions across data centers. This limitation prevents the efficient utilization of performance and resources for FLaaS. To address this, we introduce NebulaFL, which adopts a dual-layer architecture across multiple data centers in a multi-cloud environment to facilitate FL and deliver robust inference services.

To the best of our knowledge, NebulaFL is the first JointCloud-based comprehensive FL framework, which attempts to achieve effective and efficient privacy-preserving collaborative model training among data centers, including effective FL training strategy, efficient intra-data center resource scheduling, and inter-data center communication mechanism.

% while others are tailored to specific scenarios such as vehicular networks~\cite{liang2022semi} and drone swarms.
%针对异步联邦学习，一部分着眼于提高模型训练的效率和安全性【】，有一部分具体到特定的场景例如车联网，无人机群等【】，

%% file: approach.tex
\section{Our NebulaFL Approach}

\begin{figure*}[h]
    \centering
    \includegraphics[width=0.95\textwidth]{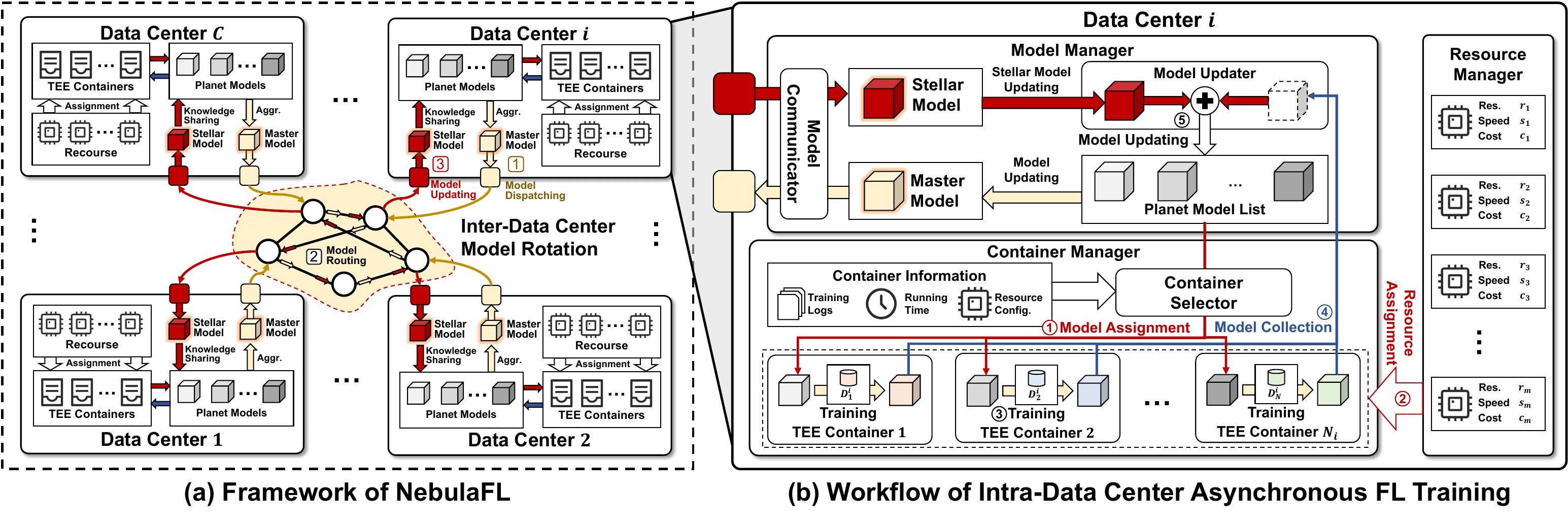}
    % \vspace{-0.2in}
    \caption{Framework and workflow of our NebulaFL approach}
    % \vspace{-0.25in}
    \label{fig:framework}
\end{figure*}

% In this section, we first outline the overall design of FedulaFL and implementation. Next, we detail the key techniques of our approach,including XXX, XXX, XXX.

%介绍下整体的，架构图，分步骤
\subsection{Overview}
Figure~\ref{fig:framework} presents the framework and workflow of our NebulaFL approach.
As shown in Figure~\ref{fig:framework} (a), NebulaFL involves multiple data centers, including multiple TEE containers and various computing resources.
NebulaFL consists of two main processes, i.e., intra-Data Center (DC) asynchronous FL training and inter-DC model rotation, respectively.

Figure~\ref{fig:framework} (b) presents the workflow of intra-DC asynchronous FL training at data center $i$.
% NebulaFL adopts multiple intermediate models, named planet models, for asynchronous FL training and periodically requests the aggregated model (i.e., master model) from other data centers as the stellar model for knowledge sharing.
% To achieve efficient FL training, 
As shown in Figure~\ref{fig:framework} (b), each data center consists of three key components, i.e., model manager, container manager, and resource manager.
The model manager maintains a list of planet models for asynchronous FL training and the stellar model for knowledge sharing.
When a planet has completed a round of training, the model manager uploads such a planet model by aggregating it with the stellar model.
In addition, the model manager periodically sends the master model aggregated by all the planet models to other data centers for knowledge sharing and updates the stellar model using the received model from other data centers.
% and requests the master model from other data centers to update the stellar model and sends the master model aggregated by all the planet models when receiving requests from other data centers.
The container manager maintains multiple TEE containers and their information, including training logs, running time, and the historical resource configuration of containers.
The container manager includes a container selector.
When a planet model is ready for local training, the container selector adaptively selects a container for local training according to the container information.
The resource manager adaptively assigns suitable hardware resources to each container according to the container information and the training information of the stellar and assigned planet model.

When the training time reaches a threshold, NebulaFL performs the inter-DC model rotation process, during which each data center sends its master model to a specific data center. 
Note that NebulaFL ensures that different data centers do not send the model to the same data center in one rotation process and each data center sends the model to different data centers in different rotation rounds.
After multiple rounds of model rotation, NebulaFL aggregates all the master models to generate a global model for deployment. 

As shown in Figure~\ref{fig:framework}, the intra-DC FL training and the inter-DC model rotation process are asynchronous.
% , where each round of intra-DC FL training consists of five steps, i.e., model assignment, resource assignment, model training, model collection, model updating, respectively, and each round of inter-DC model rotation includes three steps, i.e., model dispatching, model routing, and model updating, respectively.
% Note that NebulaFL adopts asynchronous FL training scheme, where 
In addition, each planet model is trained asynchronously without waiting for the training of other planet models.
The details of the intra-DC FL training are as follows:
\begin{itemize}
    \item \textbf{Step 1 (Model Assignment):} 
    % When a planet model is ready for training, the model manager sends the model to the container manager. 
    The container manager selects an idle container for model training based on the version (i.e., training times) of the target model and the historical training logs of the container.
    % first accesses the version (i.e., training times) of the target model and the average version of all the planet models according to the training log and selects an idle container based on the version of the model and the historical running time of the container for the model training.
    \item \textbf{Step 2 (Resource Assignment):} The resource manager assigns the most cost-effective hardware resource to the container based on the model version and the historical running time of the container.
    \item \textbf{Step 3 (Model Training):} The container uses its data to train the received planet model.
    \item \textbf{Step 4 (Model Collection):} The container sends the trained planet model to the model manager, and the container manager updates the container information.
    \item \textbf{Step 5 (Model Updating):} The model manager aggregates the trained planet model and the stellar model with a specific weight and updates the corresponding planet model in the model list.
    % Then the model manager starts a new training round for the updated planet model.
\end{itemize}
The intra-DC asynchronous training process continues until the training time threshold is reached.
Note that the stellar model updating and master model updating operations are asynchronous with the intra-DC training process.
Such two operations are only performed in the inter-DC model rotation process.
The details of the inter-DC model rotation are as follows:
\begin{itemize}
    \item \textbf{Step 1 (Model Dispatching):} When the intra-DC training time reaches a threshold, in each data center, the model manager aggregates all the planet models to generate a master model and sends the model together with the average training times of all the planet models to the target data centers. Each data center will take turns selecting a target center in different rotation rounds.
    \item \textbf{Step 2 (Model Routing):} The output port transmits the model to the target data center according to the current route table. Note that the port selects the shortest path for model transmission.
    \item \textbf{Step 3 (Model Updating):} When the input port receives the master model, the model manager uses the model to replace the old stellar model.
\end{itemize}
Note that before the model dispatching step, the model manager resets the timer to wait for the next round of model rotation. In this way, NebulaFL can periodically perform model rotation.

\begin{algorithm}[htbp]
\caption{Implementation of NebulaFL}\label{alg1}
\footnotesize
\SetAlgoLined
\SetKwFunction{Time}{Time}
\SetKwFunction{ModelAggr}{ModelAggr}
\SetKwFunction{MasterGen}{MasterGen}
\SetKwFunction{Transfer}{Transfer}
\SetKwFunction{NULL}{NULL}
\SetKwFunction{GetModel}{GetModel}
\SetKwFunction{ModelAggr}{ModelAggr}
\SetKwFunction{LocalTrain}{LocalTrain}
\SetKwFunction{CtSel}{CtSel}
\SetKwFunction{ResAssign}{ResAssign}
\SetKwFunction{PlanetUpdate}{PlanetUpdate}
\SetKwFunction{Buffer}{Buffer}
\SetKwFunction{Avg}{Avg}
\SetKwFunction{PSc}{PSc}
\SetKwFunction{Cnt}{Cnt}
\SetKwFunction{Pr}{Pr}
\SetKwFunction{RTime}{RTime}
\KwIn{\textbf{i)} $T$: training time; \textbf{ii)} $R$: available resource; \textbf{iii)} $C$: set of centers; \textbf{vi)} $CT$: set of containers for data owners; \textbf{v)} $t_c$: time interval between model rotations}%输入参数
\KwOut{ $m_g$: the global model}%输出
$T_{0} \gets \Time()$\; \label{line:time_start}
\textcolor{blue}{/* parallel  for*/}\\ \label{line:nebula_start}
\For{$i \gets 1$, $2$, ..., $|C|$}{
    \textcolor{blue}{/* initialize */} \\
    $K_i\gets |CT_i|$\; \label{line:init_start} \label{line:center_start}
    $m^i_1, m^i_2, ..., m^i_{K_i}\gets m_0$\;
    $S_i \gets [m^i_1, m^i_2, \dots, m^i_{K_i}]$ \;
    $v^i_1,...,v^i_{K_i}  \gets 0$\;
    $V_i\gets \{v^i_1,...,v^i_{K_i}\}$\;
    $c^i_{r} \gets 0$\;
    $m^i_s \gets m_0$\; \label{line:init_end}
    \textcolor{blue}{/* parallel - master model dispatching */}\\
    \While{ $\Time() - T_{0} < T$}{\label{line:rotate_dispatch_start} \label{line:rotate_start}
        \If{$\Time() - T_{0} > c^i_r\times t_c$}{\label{line:rotate_time}
            $m_i\gets \MasterGen(S_i,V_i)$\;\label{line:master_gen}
            $v_i\gets \Avg(V_i)$\;\label{line:avg_ver_gen}
            $OP_i:\Transfer(m_i, v_i, (i + c^i_r) \% |C| + 1)$\; \label{line:transfer}
            $c^i_r\gets c^i_r + 1$\;\label{line:rotate_count_update}
        }
    }\label{line:rotate_dispatch_end}
    \textcolor{blue}{/* parallel - stellar model updating */}\\
    \While{ $\Time() - T_{0} < T$}{\label{line:rotate_receive_start}
        \If{$IP_i:\Buffer() \neq \NULL$}{\label{line:IP_monitor}
            $m^i_s, v^i_s\gets IP_i:\GetModel()$\;\label{line:stellar_model_gen}
        }
    }\label{line:rotate_receive_end} \label{line:rotate_end}
    \textcolor{blue}{/* parallel  for -  planet model training */}\\
    \For{$j \gets 1$, $2$, ..., $K$}{\label{line:intra_training_start}
        % \textcolor{blue}{/* planet model training*/}\\
        \While{ $\Time() - T_{0} < T$}{
            $ct\gets \CtSel(m^i_j,v^i_j,v_s, V_i, CT_i)$\; \label{line:planet_training_start} \label{line:container_selection}
            $res \gets \ResAssign(v^i_j, v_s, V_i, ct, R_i)$\; \label{line:resource_assign}
            $m^i_j \gets ct:\LocalTrain(m^i_j,res)$\;\label{line:local_training}
            $m^i_j \gets \PlanetUpdate(m^i_j,m^i_s,v^i_j,v^i_s)$\;\label{line:planet_update}
            $v^i_j \gets v^i_j + 1$\;\label{line:version_update} \label{line:planet_training_end}
        }
 
    }\label{line:intra_training_end}
} \label{line:center_end}
$m_g \gets \ModelAggr(m_1,...,m_{|C|})$\;  \label{line:aggr}
\Return $m_g$\\ \label{line:nebula_end} 

\end{algorithm}

\subsection{Implementation of NebulaFL}
Algorithm~\ref{alg1} presents the implementation of our NebulaFL approach. 
We assume that there are $|C|$ data centers involved in FL training.
For the $i^{th}$ data center in $C$, $CT_i$ denotes the set of its TEE containers, and there are $K_i$ containers activated for local training at the same time.
Line~\ref{line:time_start} initializes the system time $T_{0}$. 
Lines \ref{line:nebula_start}-\ref{line:nebula_end} present the whole Nebula FL training process, where the ``for'' loop is a parallel loop. 
Lines \ref{line:center_start}-\ref{line:center_end} present the FL training process in each data center.
Lines~\ref{line:init_start}-\ref{line:init_end} initialize the set of containers $CT_i$, number of containers $K_i$, set of planet models $S_i$, the set of model versions $V_i$, the count of rotation rounds $c^i_r$, and the stellar model $m_s^i$, respectively.
% Line 3 and Line 4 initialize a set of planet models $S_{i}$ prepared for local training within $i$th data center. Line 6 and Line 7 initializes a list $V_{i}$, which records the version information of all the planet models. Line 8 initialize the rotation time of the master model as zero. And the stellar model is initialized as $m$ in Line 9.
Lines~\ref{line:rotate_start}-\ref{line:rotate_end} present the inter-DC model rotation process, where Lines~\ref{line:rotate_dispatch_start}-\ref{line:rotate_dispatch_end} denotes the model dispatching step and Lines~\ref{line:rotate_receive_start}-\ref{line:rotate_receive_end} indicate the stellar model updating step.
When the current training time satisfies the rotation condition, in Lines~\ref{line:master_gen} and \ref{line:avg_ver_gen}, the data center uses the function $\MasterGen(\cdot)$ to generate the master model according to all its planet models and the corresponding model versions and then calculates the average version of its planet models.
In Line~\ref{line:transfer}, the data center uses the output port $OP_i$ to transfer the master model and average version to the $(i+c^i_r)\%|C|^{th}$ data center.
In Line~\ref{line:IP_monitor}, the input port $IP_i$ monitors the input buffer.
When receiving a model, in Line~\ref{line:stellar_model_gen}, the data center uses the received model to update the stellar model with the corresponding version.
Lines~\ref{line:intra_training_start}-\ref{line:intra_training_end} show the intra-DC asynchronous FL training process.
Lines~\ref{line:planet_training_start}-\ref{line:planet_training_end} present the details of a planet model training round.
Line~\ref{line:container_selection} uses the function $\CtSel(\cdot)$ to select a container $ct$ form $CT_i$ for the training of $m_j^i$.
Line~\ref{line:resource_assign} uses the function $\ResAssign(\cdot)$ to assign a hardware resource to the selected container for model training.
Line~\ref{line:local_training} performs local training process and Line~\ref{line:planet_update} updates the planet model using the trained model and the stellar model $m_s^i$.
When the FL training process is completed, Line~\ref{line:aggr} aggregates all the master models from data centers to generate a global model for deployment.

% Specifically, the function $\StellarGen(\cdot)$ aggregates the local planet models and generates the master model $m_{i}$ at line 13. Line 14 calculates the average version of all planet models as the version of the master model $v_{i}$. Fuction $\Transfer(\cdot)$ transfers the model out of the port, and increases the number of rotations by 1 at line 16. Line 18-20 presents the step pf stellar model updating. When the Input Port Buffer of the data center is not empty, meaning there is a new staler model that needs updating, update the staler model $m_{s}$ and the version $v_{s}$.
% Line 22-28 introduces the step of $K$ planet model training. In Line 24, the fuction $\CtSel(\cdot)$ selects an idle container $ct$ based on the
% version of plane models. And the the fuction $\ResAssign(\cdot)$ assigns the most cost-effective hardware resource $rs$ to the container based on the model version and the historical running time of the container. The planet model performs local training in Line 26. The fuction $\PlanetUpdate(\cdot)$ performs planet model updates by weighted aggregation of the model with the stellar model. After updating the planet model, Line 28 updates its version. After multiple rounds
% of training, NebulaFL aggregates all the master models
% to generate a global model$m_{g}$ for deployment in Line 29. Since the global model does not participate in the actual training process, its generation can be performed asynchronously at any point in time. The following will detail the key parts of NebulaFL and analyze its convergence.

%基于版本控制的异步云际联邦训练策略 如何聚合
\subsubsection{Intra-DC Asynchronous Model Training} 
To alleviate the low training performance caused by stragglers, NebulaFL adopts an asynchronous training scheme.
Inspired by~\cite{gitfl2023}, NebulaFL uses multiple intermediate models, named planet models, rather than a global model for local training and generates the master model according to model versions, where a model with a higher version can be assigned a higher aggregation weight.

\textbf{Planet Model Updating ($\PlanetUpdate(\cdot)$)}. 
To achieve knowledge sharing among data centers, NebulaFL periodically requests the master model from other data centers as the stellar model.
Different from GitFL, which only uses the master model to update the intermediate models, NebulaFL uses the stellar model to update the planet model.
Specifically, we adopt a weighted aggregation strategy to achieve knowledge sharing from the stellar model to the planet model.
Since aggregating the straggler model results in performance degradation, we assign a higher aggregation weight to the model with a higher version.
The planet model updating operation can be defined as follows:
\begin{equation}
% \footnotesize
\begin{split}
\PlanetUpdate&(m^i_j,m^i_s,v^i_j,v^i_s) = \frac{\  w^i_{j} m_{j}^{i} + d_t m_s^{i} }{ w^i_{j} + d_t}\\
\text{s.t., }  w^i_{j} &= \max(v^i_j - v^i_s,5),
\end{split}
\end{equation}
where $w^i_{j}$ indicates the aggregation weight of the target planet model $m^i_j$ and $d_t\in[0.5,1.0]$ is a decaying weight that decreases over time and is reset to $1.0$ when model rotation.
Note that since NebulaFL updates the stellar model with a time interval rather than run-time, the stellar model becomes outdated as the training progresses.
To avoid performance degradation due to an outdated stellar model, here we assign $d_t\in[0.5,1.0]$ a decaying weight to the stellar model, where $d_t$ decreases over time and is reset to $1.0$ when model rotation.
In addition, to avoid excessive knowledge exchange, we limit the minimum value of $w^i_j$.

\textbf{Master Model Generation ($\MasterGen(\cdot)$)}.
NebulaFL periodically uses the master model of different data centers to update its stellar model, where the master model is aggregated by all the planet models, as follows:
\begin{equation}
% \footnotesize
\MasterGen(S_i,V_i) = \frac{\sum_{k=1}^{K} m_{k}^{i} \times v_{k}^{i}}{\sum_{k=1}^{K} v_{k}^{i}},
\end{equation}
where $S_i$ denotes the list of planet models in data center $i$, $K$ presents the number of planet models, and $V_i$ is the list of version information. Here, we adopt the version-guided model aggregation strategy for stellar model generation, which is still used in \cite{gitfl2023}.

% We adopt a version-guided aggregation strategy to generate the master model.

% NebulaFL generates the stellar model by merging all planet models in the data center. Since the planet model with a lower version may reduce the performance of the stellar model, NebulaFL assigns each planet model a weight based on its version to guide the merging. The stellar model generation process is shown as follows:
% \begin{equation}
% StellarGen(S_i,V_i) = \frac{\sum_{k=1}^{K} m_{k}^{i} \times v_{k}^{i}}{\sum_{k=1}^{K} v_{k}^{i}}
% \end{equation}
% where $S_i$ denotes the list of planet models in data ceter $i$, $K$ presents the number of planet models, and $V_i$ is the list of version information. At the beginning of training, planet models with higher versions should dominate the stellar model generation process as they are more accurate. At this stage, using either model versions or version differences as weights is both reasonable. However, as federated learning training progresses and all models are well-trained, we expect all planet models to participate fully in the generation process. Compared to version differences, using model versions is more likely to merge the planet models equitably.

\subsubsection{Inter-DC Model Rotation}
To achieve an efficient model communication mechanism among data centers, NebulaFL adopts the model rotation strategy rather than the traditional model aggregation strategy.
Specifically, in each model rotation round, each data center only sends its master model to one data center without waiting for the aggregation model.
In this way, compared to the traditional aggregation strategy, NebulaFL can reduce 50\% of communication overhead in each communication round.
To achieve the diversity of knowledge sharing, each data center sends the master model to different data centers in different rotation rounds.
Specifically, assume that there are $|C|$ data centers, in the $r^{th}$ rotation rounds, the $i^{th}$ data center sends its master model to the $(i+r)\% |C| + 1^{th}$ data center.
Note that when $r \% |C|-1 = 0$, each data center directly uses its master model to update the stellar model.

\subsubsection{Container Selection and Resource Scheduling Strategy}
To achieve efficient and effective FL training, NebulaFL adopts a reward-guided container selection and resource scheduling strategy.

\textbf{Container Selection ($\CtSel(\cdot)$)}.
In NebulaFL, the goal of container selection is to alleviate stragglers and ensure diversity of training.
Inspired by \cite{gitfl2023}, we aim to select the container with fewer training times for the low version model and vice versa.
However, due to i) the lack of version information, ii) training logs from other data centers, and iii) the training time of containers is based on the assigned hardware resource, the client selection strategy in \cite{gitfl2023} cannot be adapted to NebulaFL.
To achieve container selection, NebulaFL records the training logs of each container.
Since the training performance of each container is more stable compared to edge devices, based on the training logs, each data center can predict the training time of each container.
NebulaFL utilizes the historical training time of the container with a specific resource as the performance score of this container.
We use $\PSc(ct)$ to denote the performance score of the container $ct$.
Since whether the model version is outdated needs to refer to the versions of other data centers, to guide the selection, each data center records the average version of all the planet models together with the version of the stellar model $v_s$ as a metric.
Specifically, for the $i^{th}$ data center, it records the average version $v^i_{avg}$ when performing the model rotation and then calculates the version offset $v_{os} = v^i_{avg} - v_s$.
Assume that a planet model $m_p$ is ready for training and its version is $v_p$ and the current average version of the planet model in the $i^{th}$ data center is $v_i$.
Based on $v_{os}$, we can get
\begin{equation}\label{eq:delta_v}
    \Delta v_{p} = v_p - v_i - v_{os}.
\end{equation}

Based on $\Delta v_{p}$, the performance reward of a container $ct$ can be calculated as follows:
\begin{equation}
\footnotesize
\begin{aligned}
R_{p}(ct,m_p) = 
\begin{cases}
    \frac{\PSc(ct)}{\max_{ct^\prime \in CT_i}\PSc(ct^\prime)} \Delta v_{p}, &\Delta v_{p}> 0\\
    (1-\frac{\min_{ct^\prime \in CT_i}\PSc(ct^\prime) - \PSc(ct)}{\max_{ct^\prime \in CT_i}\PSc(ct^\prime)})\Delta v_{p}, &\Delta v_{p}\leq 0\\
\end{cases}
,
% (v_p - v_i - v_{os}) \times  \frac{ \PSc(ct) - \frac{\sum_{ct^\prime \in CT_i} \PSc(ct^\prime)}{|CT_i|}}{\max_{ct^\prime \in CT_i} \PSc(ct^\prime)}.
\end{aligned}
\end{equation}
where for $\Delta v_{p}>0$, the container $ct$ with a higher performance score let to a higher performance reward and for $\Delta v_{p}\leq 0$, the container $ct$ with a lower performance score let to a higher performance reward.

To ensure that containers can participate in training more fairly, NebulaFL adopts the curiosity reward~\cite{pathak2017curiosity,gitfl2023,hu2022accelerating} together with the performance reward to guide container selection, where a container that has been selected more times achieves a lower curiosity reward.
Assume that $\Cnt(ct)$ denotes the number of selected times for the container $ct$.
We can use Model-based Interval Estimation with Exploration Bonuses (MBIE-EB)\cite{bellemare2016unifying} to measure the curiosity reward for the container $ct$ as follows:
\begin{equation}
\footnotesize
\begin{aligned}
R_{c} (ct)=\frac{1}{\sqrt{\Cnt(ct)}}.
\end{aligned}
\end{equation}

To ensure the diversity of selection, NebulaFL selects containers according to the specific probabilities calculated by the above two rewards.
For the planet model $m_p$ in the $i^{th}$ data center, the selection probability of a container $ct$ can be calculated as follows:
\begin{equation}
\footnotesize
\begin{aligned}
P(ct)=\frac{R_p(ct,m_p)+R_c(ct)}{\sum_{ct^\prime \in CT_i} (R_p(ct^\prime,m_p)+R_c(ct^\prime))}.
\end{aligned}
\end{equation}

\textbf{Resource Scheduling ($\ResAssign(\cdot)$)}.  
NebulaFL aims to assign the most cost-effective resource to the selected container, which balances training time and cost.
Specifically, NebulaFL prefers to assign low-cost resources to the container with the planet model with a higher version and assign high-performance resources to that with the planet model with a lower version.
Therefore the resource scheduling of NebulaFL is based on two factors, i.e., cost factor and time factor.
Assume that $R_i$ denotes the set of available resources of the $i^{th}$ data center and $\Pr_i(r,ct)$ denotes the unit time price of the resource $r$  for container $ct$ in the $i^{th}$ data center.
For a resource $r$ and container $ct$, its cost factor can be calculated as follows:
\begin{equation}
\footnotesize
F_{cost}(r,ct)=   \frac{\Pr_i(r,ct) - \frac{\sum_{r^\prime\in R_i} \Pr_i(r^\prime,ct)}{|R|}}{\max_{r^\prime\in R_i} \Pr_i(r^\prime,ct)},
\end{equation}
where a smaller cost factor indicates that the resource is cheaper.
Since NebulaFL prefers to assign high-performance resources to containers with low version model, we calculate the time factor based on $\Delta_{v_p}$ defined by Equation~\ref{eq:delta_v}, where a lower $\Delta_{v_p}$ is more sensitive to the time factor.
The time factor is calculated as follows:
\begin{equation}
\footnotesize
F_{time}(r, ct, m_p)= \frac{\RTime(r,ct) - \frac{\sum_{r^\prime \in R_i} \RTime(r^\prime,ct)}{|R|}}{\max_{r^\prime \in R_i} \RTime(r^\prime,ct)}\Delta_{v_p},
\end{equation}
where $\RTime(r,ct)$ indicates the historical runing time of container $ct$ with the resource $r$.
NebulaFL assigns the resource with the minimized value of the sum of two factors to the target container.

\subsection{Convergence Analysis}
Inspired by the work presented in paper\cite{Li2020On}, which demonstrates the convergence of FedAvg\cite{fedavg2019} under certain conditions on Non-IID data: i) the local objective $f_k(w)$ is $L$-smooth, ii) $f_k(w)$ is $\mu$-strongly convex, iii) during each training round, a subset of $m$ containers is randomly selected to participate with probabilities $p_1,...,p_N$, and the model aggregation is a simple average. NebulaFL is a variant based on FedAvg; therefore, the convergence proof process is similar.

\textbf{Notation.}
Assume that all containers adopt Stochastic Gradient Descent (SGD) as the optimizer.
Let $w^k_t$ represent the parameters of the $k$-th middleware model at the $t$-th SGD iteration. Let $cr$ represent the number of global rotations. Define $g^k_t$ as the gradient obtained by training the model on the selected container with a data batch $\xi^k_t$, i.e., $g^k_t = \nabla f_k(w^k_t, \xi^k_t)$ Then, the update of NebulaFL can be described as follows:
\begin{equation}
\footnotesize
v^k_t = w^k_t - \eta_t g^k_t,
\end{equation}
\begin{equation}
\footnotesize
w^k_{t+1} = \epsilon v^k_t+(1-\epsilon) \overline{w}_{cr-1}^{ci-1},
\end{equation}
where $\epsilon \in [0,1]$ and $\overline{w}_{cr-1}^{ci-1}$ be the aggregated model (Stellar Model) rotated from the previous data center. Thus, $\overline{w}_{cr}^{ci}$ or simply $\overline{w}_{t}$ and $\overline{v}_t$ can be expressed as:
\begin{equation}
\footnotesize
\overline{w}_{t} = \frac1m \sum_{k=1}^{m} w^k_t,\quad \overline{v}_t = \frac1m \sum_{k=1}^{m}v^k_t,
\end{equation}

\textbf{Assumptions and Lemmas.}
Inspired by \cite{Li2020On,lin2018don,yu2019parallel,zhang2012communication}, the convergence of NebulaFL is based on the following four assumptions:
% Based on the Assumptions and Lemmas in \cite{Li2020On}, we have made the necessary modifications to the relevant parts to account for the changes introduced by NebulaFL. We present the following Assumptions and Lemmas: 

\noindent \textbf{Assumption 1.}  $f_1, \dots , f_N$ \textit{are all} $L$\textit{-smooth: for all} $v$ \textit{and} $w$, $f_k(v)\le f_k(w) + (v-w)^T \nabla f_k(w) +\frac{L}{2} \left \|  v-w \right \|^2_2 $.

\noindent \textbf{Assumption 2.} $f_1, \dots ,f_N$ \textit{are all} $\mu$\textit{-convex: for all} $v$ \textit{and} $w$, $\left \| \nabla f_k(w)-\nabla f_k(v) \right \|_2 \ge \mu \left \| w-v \right \|_2$.

\noindent \textbf{Assumption 3.} \textit{The variance of stochastic gradients in each container and the expectation of squared norm of stochastic gradients are bounded:} $\mathbb{E} \left \| \nabla f_k(w^k_t ,\xi^k_y) - \nabla f_k(w^k_t) \right \| ^2 \le \sigma^2_k,\ \mathbb{E}\left \| \nabla f_k(w^k_t ,\xi^k_y) \right \|^2\le G^2 $ \textit{for all} $k = 1, \dots , N$\textit{, where} $\xi^k_t$ \textit{be sampled from the} $k$\textit{-th container's local data uniformly at random.} 

\noindent \textbf{Assumption 4.} \textit{Assume} $\mathcal{S}_t$ \textit{is the set of indices of } $m$ \textit{containers, which are randomly selected with replacement according to probabilities} $p_1,\dots,p_N$. \textit{The aggregation step of NebulaFL performs} $w_t \gets  \frac1m \sum_{k\in \mathcal{S}_t} w^k_t$.

\noindent Based on Assumptions 1, 2, 3, and 4, we can derive Lemmas 1, 2, and 3 as stated in \cite{Li2020On}. However, Lemma 4(Unbiasedness) has a slight difference. Since Nebula aggregates the model with the Stellar model before broadcasting, the expected value of the model parameters $\mathbb{E} \overline{w}_t$ is less than the weighted average of all trained model parameters. Specifically, $\mathbb{E} \overline{w}_t = \frac{\epsilon m}{m+\epsilon -1 } \overline{v}_t \le \overline{v}_t $, where $\epsilon \in [0,1]$. This is anticipated, as our method converges faster than FedAvg\cite{fedavg2019}. Consequently, $\overline{w}_t$ is not an unbiased estimate of $\overline{v}_t$, allowing us to achieve a smaller convergence bound compared to \cite{Li2020On}.

\textbf{Theorem 1.} \textit{Let} $L,\ \mu,\ \sigma^k,\ G$ \textit{be defined therein.} $E$ \textit{be the number of SGD iterations in one round of FL training, and let} $cr$ \textit{be the number of rotations in Nebula. Thus,} $t = cr \times E$ \textit{is the total number of SGD iterations up to the} $cr$\textit{-th round. Choose} $\gamma = \max\{\frac{8L}{\mu},E\}-1$ \textit{, and the learning rate} $\eta_t = \frac{2}{\mu (t+\gamma)}$, \textit{then we can derive the following:} 
% \begin{strip}
\begin{equation}
\footnotesize
\scalebox{0.95}{$
    \mathbb{E} \left [  F(\overline{w})\right ] -F^\star \le \frac{L}{2\mu(t+\gamma )} \left [ \Delta + (\gamma  +1 ) \mu  \left \| \overline{w}_0 -w^\star  \right \|^2_2  \right ],
$}
\end{equation}
% \end{strip}
\noindent \textit{where} $\Delta = \frac{4B+\frac{16}{m}E^2G^2 }{\mu}$, \textit{and} $B$ \textit{is defined in} \cite{Li2020On}.
% $B = \sum _ { k = 1 } ^ { N } p _ { k } ^ { 2 } \sigma _ { k } ^ { 2 } + 6 L \Gamma + 8 ( E - 1 ) ^ { 2 } G ^ { 2 }$. $\Gamma$\textit{is the quantification of Non-IID degree.}

%% file: experiment.tex
%%第一张大表
\input{B1table20WNoIID}
\section{Performance Evaluation}

%介绍实验环境，以及整个章节的安排

\subsection{Experimental Setup}
To demonstrate the effectiveness of our NebulaFL, we conducted experiments on five baselines using three models on three datasets.
All the experiments were conducted on one Ubuntu workstation with 22 vCPU, 90GB memory, and an NVIDIA RTX 4090 GPU.

\textbf{Settings of Data Centers.}
We consider a realistic multi-cloud (i.e., multiple data centers) scenario and have investigated the data transfer bandwidth between different data centers and the heterogeneous resource prices at each center, such as NVIDIA A100, NVIDIA RTX 4090, and 3080. We refer to the real performance and price of various resources of the computing resource platform\cite{AutoDL} to set configurations. To more accurately reflect the performance differences between homogeneous resources caused by other factors in real-world scenarios, we simulate the runtime of each resource based on a normal distribution.

\textbf{Datasets and models.}
We compared the performance of all approaches on three well-known datasets, i.e., CIFAR-10~\cite{krizhevsky2009learning}, CIFAR-100~\cite{krizhevsky2009learning}, and Fashion-MINIST\cite{xiao2017fashion}. To evaluate the performance in Non-IID scenarios, we adopt the Dirichlet distribution~\cite{hsu2019measuring} $Dir(\alpha)$
% , to modulate the data heterogeneity $\alpha$ across data owners using
to divide the CIFAR-10 and CIFAR-100 datasets. Here, we set $\alpha$ as to 0.1, 0.5, and 1.0, with a smaller value corresponding to higher data heterogeneity among containers. We did not apply to the Fashion-MINIST dataset, as it inherently exhibits Non-IID characteristics. We selected multiple popular models to verify the effectiveness of the method, including CNN, ResNet-18, and VGG-16.
% We use the variable $\alpha$ to represent the degree of heterogeneity of datasets, following the Dirichlet distribution~\cite{hsu2019measuring}. 

\textbf{Baselines.} 
% We compare the performance of NebulaFL with the following five baselines. 
To comprehensively evaluate our method, we considered the classic FL method FedAvg~\cite{fedavg2019} as well as four state-of-the-art methods, i.e.,  FedHiSyn~\cite{li2022fedhisyn}, FedHKT~\cite{deng2023hierarchical}, FedSA~\cite{ma2021fedsa} and Pisces~\cite{jiang2022pisces}. The first three are synchronous training frameworks, while the last two are asynchronous frameworks.
% , with FedSA being a semi-asynchronous training framework.
%其中 经典的FedAvg方法 以及先进的算法，包括Avg、HKT、FedHisyn、FedSA、Pieces、gitFL, 其中前三个为同步训练框架，后三个为异步框架，其中FedSA为半异步训练框架。
\begin{itemize}
    \item \textbf{FedAvg}~\cite{fedavg2019}: After several training epochs, the local models are uploaded to the server, which then calculates the average and broadcasts it back to the clients.
    \item \textbf{FedHiSyn}~\cite{li2022fedhisyn}: After clustering the clients, sort them by communication time. Upon completing the training, rotate the model to the next client.
    \item \textbf{FedHKT}~\cite{deng2023hierarchical}: Utilize a hybrid knowledge transfer mechanism combined with a weighted ensemble distillation scheme to transfer specialized knowledge.
    \item \textbf{FedSA}~\cite{ma2021fedsa}: The server aggregates the local models within a specified time window during each round.
    \item \textbf{Pisces}~\cite{jiang2022pisces}: Implement a novel scoring mechanism to identify suitable clients, thereby avoiding excessive resource costs and outdated training computations.
\end{itemize}
For baselines and NebulaFL, we adopt the SGD optimizer with a learning rate of 0.01 and a momentum of 0.9. For the hyperparameters, we set the batch size to 50 and the local training epoch to 5.

% \textbf{Parameters.}
% For baselines and NebulaFL, we adopt the SGD optimizer with a learning rate of 0.01 and a momentum of 0.9. For the hyperparameters, we set the batch size to 50 and the local traing epoch to 5. We use the variable $\alpha$ to represent the degree of heterogeneity of datasets, following the Dirichlet distribution~\cite{hsu2019measuring}. We ran all the experiments on one Ubuntu workstation with 22 vCPU, 90GB memory and a NVIDIA RTX 4090 GPU.

% \textbf{Evaluation Metrics.} 
% We mainly focus on the following three metrics: 1) The best accuracy that different methods can achieve with data heterogeneity and model heterogeneity; 2) The communication volume required to achieve the target accuracy in a wide-area data centers transmission scenario; 3) The time and cost required to achieve the target accuracy under different situations.

\subsection{Performance Comparison}
To fully demonstrate the effectiveness of NebulaFL, we compared it with five baselines on three different datasets in three Non-IID scenarios. We ran a performance comparison across multiple dimensions, including accuracy, communication overhead, training time, and cost.
\input{B2table}
\subsubsection{Comparison of Accuracy}

Table~\ref{table:1} presents the classification accuracy results for NebulaFL and all the five baselines.
% , showcasing its performance under varying conditions of model type, dataset, and degree of data heterogeneity. 
% The first two column lists the different AI models and datasets, the third column indicates the degree of data heterogeneity including 0.1, 0.5, 1.0 based on Diricht distribution~\cite{hsu2019measuring}. The remaining columns show the maximum test accuracy and standard deviation achieved by all methods within the same time frame.
From Table~\ref{table:1}, we can observe that NebulaFL achieves the highest accuracy in almost all cases. As an example of dataset CIFAR-10, when $\alpha=0.5$, NebulaFL can achieve an improvement of 5.71\% over the second-best baseline FedHiSyn. 
% In addition, NebulaFL shows significant improvement with a data heterogeneity of 0.1, indicating that it is well-suited for handling data heterogeneity. 
Although our performance is slightly below the best baseline in the VGG-16, CIFAR-100 configuration with a data heterogeneity of 0.5, NebulaFL achieves excellent results in terms of time and cost efficiency under the same conditions, which will be analyzed in Sections~\ref{sec:comm} and \ref{sec:time}. 
% We also observe that NebuLaFL achieves greater improvements on the CIFAR-10 and CIFAR-100 datasets compared to Fashion-MNIST. This is because the image classification tasks on these two datasets are more challenging than on Fashion-MNIST. While all methods perform well on Fashion-MNIST, NebulaFL still outperforms all the baseline methods, achieving the best results.

\begin{figure}[htbp] 
% \vspace{-0.1in}
\begin{center} 
\includegraphics[width=0.9\linewidth]{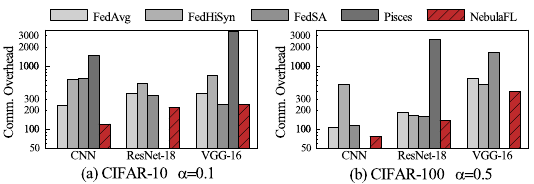}
\end{center}
    % \vspace{-0.15in}
    \caption{Comparison of communication overhead with different configs}
    \label{fig2}
    % \vspace{-0.15in}
\end{figure}

\subsubsection{Comparison of Communication Overhead}\label{sec:comm}
Figure~\ref{fig2} presents the results of communication overhead for all methods in different datasets and models.
Since the network bandwidth of inter-data centers is significantly lower than that of the intra-data center, we only compared the communication overhead of inter-data centers.
% The internal network within data centers is robust and the communication cost is negligible; hence, we can ignore it. 
% However, the network bandwidth between data centers is relatively low and variable, making it crucial to compare the communication overhead differences between NebulaFL and the baseline methods. 
% Given that the model sizes are identical, we further measured the communication overhead of each method by counting the number of communications required to achieve the target training accuracy. 
Since FedHKT involves transmitting intermediate logits and is not directly comparable, we focus on comparing NebulaFL with the other four baselines. We observe that NebulaFL incurs less communication overhead than all the four methods. For instance, in the case of the CNN model on the CIFAR-10 dataset, NebulaFL reduces communication rounds by up to 50\% compared to the best baseline, FedAvg. 
% From a theoretical analysis perspective, for the FedAvg method, each training round requires the upload and download of $K$ models. Even if one center acts as the server, at least $2(K-1)$ model transmissions are needed. In contrast, NebulaFL's design involves rotating the stellar model between data centers at each rotation time, which only requires $K$ model transmissions per round. Thus, by theoretical analysis, NebulaFL results in lower communication overhead compared to the traditional FedAvg method, potentially reducing communication time by up to 50\% with the same training round.

\subsubsection{Comparison of Training Time and Cost}\label{sec:time}
Table~\ref{table2} presents the time and cost to achieve the target accuracy under six methods using the ResNet-18 model on the CIFAR-10 dataset. 
From this table, we can find that, from the perspective of training time, NebulaFL outperforms all the baselines, being 9.96\% to 63.41\% faster than the best baseline. From the perspective of cost, NebulaFL outperforms all the baselines in five out of six cases, with a maximum cost saving of 61.94\% compared to the best baseline. We can observe when the accuracy target becomes higher, our approach uses much less training time and cost than the other baselines. As an example of $\alpha$ = 0.1 and the target accuracy is 70\%, compared to the best baseline FedHiSyn, NebulaFL saved 55.82\% in cost and achieved a 63.41\% faster training time. This indicates that our method offers better time and cost efficiency at the same level of accuracy.

\subsection{Scalability Analysis}
To demonstrate the versatility and scalability of the NebulaFL method in different scenarios, we examined the impact of various configurations on NebulaFL from the following three perspectives: the number of data centers, the total number of containers, and the number of active containers.

\begin{figure}[htbp]
% \vspace{-0.1in}
    \centering
    \includegraphics[width=\linewidth]{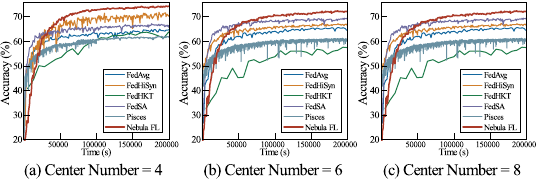}
    % \vspace{-0.25in}
    \caption{Learning curves for different numbers of centers}
    \label{figC1}
    % \vspace{-0.2in}
\end{figure}

\subsubsection{Impacts of Different Numbers of Data Centers}
To verify the effectiveness of the method under different data center configurations, we set up three different configurations, including the number of data centers and resource configurations. The number of data centers was set to $4$, $6$, and $8$, with resource conditions randomly generated based on real data surveyed from the AutoDL platform. We tested using the ResNet-18 model on the CIFAR-10 dataset, and the results are shown in Figure~\ref{figC1}. From this learning curve figure, it can be seen NebulaFL achieved the highest accuracy across all settings. Although its early convergence speed was slower than other baselines, it quickly surpassed them and achieved the highest accuracy. This indicates that as the number of data centers increases, NebulaFL's performance remains unaffected, demonstrating the scalability and general applicability of the architecture.

\begin{figure}[htbp]
% \vspace{-0.1in}
    \centering
    \includegraphics[width=0.9\linewidth]{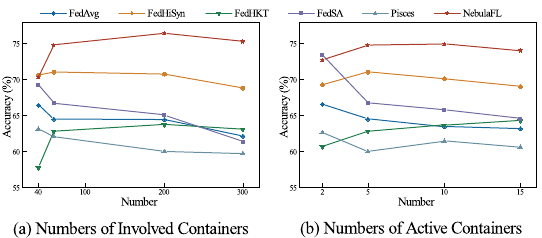}
    % \vspace{-0.15in}
    \caption{Learning curves for different numbers of centers}
    \label{figC2}
    % \vspace{-0.15in}
\end{figure}

\subsubsection{Impacts of Different Numbers of Involved Containers}
To test the impact of the number of containers on performance, we conducted experiments with four different settings of the number of involved devices, i.e., $|C|$ = 40, 100, 200, and 300, respectively, on Cifar10 dataset using ResNet-18 model, shown as the Fig.~\ref{figC2}(a). From the table, we can find that NebulaFL achieves the highest accuracy with different numbers of involved containers. when more containers are involved in the training process, the more improvements NebulaFL can obtain compared with baseline methods. This is because, as the number of participating containers increases, NebulaFL has a broader range of container selections. This allows NebulaFL to effectively combine and share knowledge from different containers, thereby achieving better accuracy.

%这是因为当参与的container越多时，NebulaFL的容器选择性范围更广，NebulaFL可以充分结合不同container进行知识共享从而达到一个较好的精度。

\subsubsection{Impacts of Different Numbers of Active Containers}
To investigate the impacts of the number of simultaneously training containers on NebulaFL, we considered four different activated container settings, where the numbers of simultaneously training containers are 2, 5, 10, and 15 within a data center, respectively. Fig.~\ref{figC2}(b) shows that NebulaFL maintains strong performance as the number of simultaneously trained containers increases. In such scenarios, NebulaFL's accuracy shows a more significant advantage over the baselines, indicating its robustness in multi-container training environments. Among the baselines, FedHKT also improves as the number of containers increases; however, the overall trend for the baselines is a decline. This decline can be attributed to the excessive number of containers simultaneously training, which can lead to network congestion and increased latency, thereby hindering performance. Additionally, the increased variability in data distribution among more containers can exacerbate the challenges of data heterogeneity, further impacting the effectiveness of model training in baseline methods.

%实验D 消融实验
\subsection{Ablation Study}
Next, we conducted ablation experiments on two key designs in NebulaFL to verify their effectiveness: i) container selection and resource scheduling strategy, ii) model rotation strategy and different rotation time.

\input{D1Table}

\subsubsection{Container Selection and Resource Scheduling Strategy}
To demonstrate the validity of our proposed container selection and resource scheduling strategy, we developed five variants of NebulaFL: i) ``Random'' that selects container and resource randomly; ii) ``NebulaFL w RR'' that selects random resource; iii) ``NebulaFL w RC'' that selects random container; iv) ``NebulaFL wo C'' that selects resource without cost factor;  v) ``NebulaFL wo T'' that selects resource without time factor. Table ~\ref{D1Table} presents the ablation study results on CIFAR-10 dataset with ResNet-18 following Non-IID distribution with 1.0. We can observe that NebulaFL achieved the shortest time and lowest cost to reach the same accuracy across the six design methods, indicating that our container selection and resource scheduling strategy is effective. The ``NebulaFL wo C'' variant demonstrated the shortest time, validating the effectiveness of our time factor. However, the ``NebulaFL wo T'' variant, which relies on the cost factor for greedy selection, did not yield the lowest cost. This outcome suggests that the focus on minimizing costs led to increased instances of stragglers, thereby prolonging the training time and increasing overall costs. Thus, considering both time and cost factors concurrently proves to be an effective approach.
%我们可以观察到NebulaFL在六种设计中在达到相同精度时获得了最少的时间和成本，这说明我们设计的容器选择与资源调度策略是有效的。``NebulaFL wo C''在时间是用时最小的，说明我们的time factor因子是有效的。然而NebulaFL wo T依据于成本因子进行贪心选择，成本却不是最低的，原因在于贪心成本导致掉队者问题加剧，从而训练时间增长，从而成本增加，因此我们同时考虑时间与成本两个因子是有效的。

\begin{figure}[htbp]
% \vspace{-0.1in}
    \centering
    \includegraphics[width=0.9\linewidth]{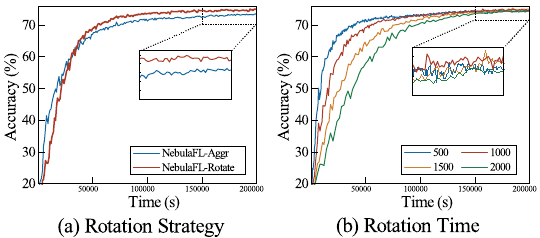}
    % \vspace{-0.15in}
    \caption{Ablation study aboout rotation strategy}
    \label{figD2}
    % \vspace{-0.15in}
\end{figure}

\subsubsection{Rotation Strategy Ablation}
Considering the communication overhead between data centers, we adopted a periodic rotation strategy to reduce the number of communications while ensuring knowledge sharing between different data centers. To verify the effectiveness of the rotation strategy, we applied the classic synchronous aggregation method from FedAvg\cite{fedavg2019} for model aggregation between data centers while keeping the internal training approach within the data centers unchanged, as a new baseline NebulaFL\-Aggr.  
Fig.~\ref{figD2}(a) shows the learning curves of NebulaFL\-Aggr and NebulaFL\-Rotate using the ResNet-18 model on the CIFAR-10 dataset. We can observe that NebulaFL\-Rotate can achieve the best inference accuracy and consistently outperform the aggregation method. This indicates that the inter-DC model rotation strategy not only helps reduce communication overhead but also positively contributes to performance improvement.
To further explore the impact of rotation time on the learning curve, we conducted experiments by setting the rotation time to 500, 1000, 1500, and 2000 while keeping other conditions constant. As shown in Fig.~\ref{figD2}(b), a shorter rotation time results in faster early-stage convergence. This is because shorter rotation times enable faster knowledge exchange between models, but overly frequent synchronization can lead to premature convergence, thereby slightly reducing the final convergence accuracy.

%Although the final accuracy is not significantly affected by the rotation time, a rotation time of 1000 slightly outperforms the other settings. This is because a shorter rotation time allows for quicker knowledge sharing between models, but excessively rapid synchronization can lead to reduced final convergence accuracy. 

%% file: B1table20WNoIID.tex
\renewcommand{\arraystretch}{1.1}
\begin{table*}[htbp]
  
  \centering
  \caption{TEST ACCURACY COMPARISON FOR NON-IID SCENARIOS USING THREE DL MODELS}
  % \vspace{-0.1in}
  \label{table:1}
  \resizebox{0.9\textwidth}{!}{
    % \begin{tabular}{|c|>{\centering\arraybackslash}m{1.3cm}|c|c|c|c|c|c|c|}
    \begin{tabular}{|c|c|c|c|c|c|c|c|c|}
    \hline
    \makecell[c]{\multirow{2}{*}{Model}} &  \makecell[c]{\multirow{2}{*}{Datas.}} & Heter. & \multicolumn{6}{c|}{Test Accuracy (\%)} \\
    \cline{4-9}
     &  & Set.  & FedAvg & FedHiSyn & FedHKT & FedSA & Pisces & \textbf{NebulaFL }(Ours)\\
    \hline
    \noalign{\vskip 0.03cm}
    \hline
    \multirow{7}{*}{CNN}
    & \multirow{3}{*}{CIFAR-10}
    & 0.1 & 44.92$\pm$2.86 & 46.02$\pm$2.01 & 39.26$\pm$2.20 & \underline{50.06$\pm$1.39} & {47.8$\pm$0.53} & \textbf{52.9$\pm$0.44} \\ 
    &
    & 0.5 & \underline{55.03$\pm$0.62} & 53.12$\pm$0.95 & 43.36$\pm$1.35 & 55.03$\pm$0.71 & 52.23$\pm$0.61 & \textbf{55.68$\pm$0.28} \\ 
    &
    & 1.0 & \underline{56.61$\pm$0.36} & 56.06$\pm$1.14 & 48.94$\pm$0.85 & 56.11$\pm$0.33 & 51.95$\pm$0.22 & \textbf{57.18$\pm$0.42} \\ 
    % &
    % & \textit{IID} & 57.28$\pm$0.14 & 57.08$\pm$0.26 & 46.45$\pm$1.56 & \underline{57.52$\pm$0.14} & 57.48$\pm$0.1 & \textbf{58.69$\pm$0.14} \\ 
    \cline{2-9}
    & \multirow{3}{*}{CIFAR-100}
    & 0.1 & 29.67$\pm$0.51 & 25.7$\pm$1.66 & 24.98$\pm$0.97 & \underline{29.87$\pm$0.60} & 29.19$\pm$0.32 & \textbf{29.89$\pm$0.17} \\ 
    & 
    & 0.5 & 32.53$\pm$0.28 & 30.48$\pm$0.63 & 30.26$\pm$0.61 & \underline{33.61$\pm$0.55} & 30.84$\pm$0.22 & \textbf{34.16$\pm$0.33} \\ 
    & 
    & 1.0 & 31.63$\pm$0.24 & 32.09$\pm$0.33 & 29.21$\pm$0.71 & \underline{32.14$\pm$0.26} & 29.94$\pm$0.10 & \textbf{33.36$\pm$0.17} \\ 
    % & 
    % & \textit{IID} & 32.25$\pm$0.22 & 31.9$\pm$0.69 & 29.17$\pm$0.49 & \underline{32.43$\pm$0.2} & 32.38$\pm$0.22 & \textbf{33.97$\pm$0.14} \\ 
    \cline{2-9}
    & F-MNIST & - & 89.29$\pm$0.10 & 89.2$\pm$0.32 & 78.13$\pm$1.06 & \underline{89.38$\pm$0.10} & 88.99$\pm$0.17 & \textbf{89.74$\pm$0.14} \\
 
    \hline

    \hline
    \multirow{7}{*}{ResNet-18} 
    & \multirow{3}{*}{CIFAR-10} 
    & 0.1 & 43.42$\pm$1.46 & 47.2$\pm$4.12 & \underline{53.03$\pm$5.87} & 49.54$\pm$1.32 & 40.22$\pm$1.65 & \textbf{55.85$\pm$1.33} \\
    &
    & 0.5 & 60.24$\pm$0.42 & \underline{65.2$\pm$0.36} & 59.88$\pm$0.77 & 63.84$\pm$0.22 & 58.11$\pm$0.28 & \textbf{70.91$\pm$0.10} \\ 
    &
   & 1.0 & 64.5$\pm$0.32 & \underline{71.05$\pm$0.66} & 62.85$\pm$0.81 & 66.77$\pm$0.20 & 59.99$\pm$0.24 & \textbf{74.83$\pm$0.14} \\
    % &
    % & \textit{IID} & 64.08$\pm$0.1 & \textbf{73.25$\pm$0.39} & 63.77$\pm$0.48 & 65.0$\pm$0.14 & 63.45$\pm$0.10 & \underline{71.08$\pm$0.10} \\ 
    \cline{2-9}
    & \multirow{3}{*}{CIFAR-100}
    & 0.1 & 33.12$\pm$0.36 & 33.01$\pm$1.86 & \underline{35.87$\pm$0.68} & 35.34$\pm$0.56 & 30.06$\pm$0.39 & \textbf{39.06$\pm$0.19} \\ 
    & 
    & 0.5 & 41.05$\pm$0.32 & \underline{45.39$\pm$0.85} & 39.87$\pm$0.48 & 42.1$\pm$0.39 & 39.49$\pm$0.17 & \textbf{47.55$\pm$0.17} \\ 
    & 
    & 1.0 & 42.78$\pm$0.17 & \underline{47.86$\pm$0.87} & 39.35$\pm$1.17 & 43.35$\pm$0.22 & 41.03$\pm$0.14 & \textbf{49.47$\pm$0.14} \\
    % & 
    % & \textit{IID} & 41.91$\pm$0.17 & \textbf{49.64$\pm$0.49} & 40.86$\pm$0.41 & 42.69$\pm$0.17 & 41.84$\pm$0.10 & \underline{47.01$\pm$0.14} \\
    \cline{2-9}
    & F-MNIST & -  & \underline{90.76$\pm$0.10} & 90.66$\pm$0.17 & 86.38$\pm$0.65 & 90.44$\pm$0.03 & 90.61$\pm$0.10 & \textbf{91.29$\pm$0.10} \\ 
    \hline

    \hline
    \multirow{7}{*}{VGG-16} 
    & \multirow{3}{*}{CIFAR-10} 
    & 0.1 & 60.22$\pm$3.66 & 63.23$\pm$4.95 & \underline{68.59$\pm$2.81} & 67.33$\pm$3.41 & 64.62$\pm$0.72 & \textbf{72.66$\pm$0.64} \\ 
    &
    & 0.5 & 77.51$\pm$0.69 & \underline{79.73$\pm$0.26} & 73.86$\pm$1.32 & 78.79$\pm$0.22 & 76.03$\pm$0.14 & \textbf{81.20$\pm$0.30} \\ 
    &
    & 1.0 & 78.93$\pm$0.69 & \underline{81.52$\pm$0.55} & 75.85$\pm$0.57 & 80.10$\pm$0.35 & 77.67$\pm$0.00 & \textbf{83.55$\pm$0.22} \\ 
    % &
    % & \textit{IID} & 79.84$\pm$0.00 & \underline{84.02$\pm$0.54} & 77.15$\pm$1.35 & 80.14$\pm$0.00 & 78.93$\pm$0.00 & \textbf{84.52$\pm$0.00} \\ 
    \cline{2-9}
    & \multirow{3}{*}{CIFAR-100}
    & 0.1 & \underline{47.00$\pm$0.85} & 42.99$\pm$2.42 & 45.65$\pm$1.89 & \textbf{47.92$\pm$0.77} & 46.30$\pm$0.33 & 46.83$\pm$0.34 \\ 
    & 
    & 0.5 & 54.27$\pm$0.30 & \underline{56.08$\pm$0.77} & 49.16$\pm$1.81 & 54.18$\pm$0.85 & 53.49$\pm$0.10 & \textbf{58.59$\pm$0.33} \\ 
    & 
    & 1.0 & 55.52$\pm$0.35 & \underline{59.32$\pm$0.57} & 49.20$\pm$2.26 & 55.26$\pm$0.40 & 54.30$\pm$0.10 & \textbf{60.28$\pm$0.24} \\ 
    % & 
    % & \textit{IID} & 56.78$\pm$0.24 & \underline{62.73$\pm$0.46} & 50.36$\pm$1.38 & 57.70$\pm$0.20 & 55.85$\pm$0.10 & \textbf{63.18$\pm$0.00} \\ 
    \cline{2-9}
    & F-MNIST & -  & 92.99$\pm$0.10 & \underline{93.20$\pm$0.00} & 88.27$\pm$1.21  & \underline{93.20$\pm$0.00} & 92.23$\pm$0.10 & \textbf{93.38$\pm$0.10} \\ 
    \hline

    % \hline
    % LSTM & ShakeSpeare & - & \textbf{52.83$\pm$0.14} & 49.84$\pm$1.64 & 46.81$\pm$0.45 & \underline{49.91$\pm$0.47} & 49.76$\pm$0.10 & 48.05$\pm$0.17 \\ 

    % \hline
    
    \end{tabular}%
}
% \vspace{-0.2in}
\end{table*}%

%% file: B2table.tex
\begin{table*}[htbp]
  \centering
  \caption{COMPARISON OF TRAINING TIME AND COST TO ACHIEVE GIVEN TARGET INFERENCE ACCURACY}
  % \vspace{-0.1in}
  \label{table2}
  \resizebox{0.9\textwidth}{!}{
    \begin{tabular}{|c|c|c|c|c|c|c|c|c|c|c|c|c|c|}
    \hline
    Heter. & \multirow{2}{*}{Acc. (\%)} & \multicolumn{2}{c|}{FedAvg} & \multicolumn{2}{c|}{FedHiSyn} & \multicolumn{2}{c|}{FedHKT} & \multicolumn{2}{c|}{FedSA} & \multicolumn{2}{c|}{Pisces} & \multicolumn{2}{c|}{\textbf{NebulaFL} (Ours)} \\
    \cline{3-14}
     Set. & & Time (h) & Cost (\$) & Time (h) & Cost (\$) & Time (h) & Cost (\$) & Time (h) & Cost (\$) & Time (h) & Cost (\$) & Time (h) & Cost (\$) \\
    \hline
    
    \noalign{\vskip 0.03cm}
    
    \hline
    \multirow{2}{*}{0.1} 
    & 47 & -&-&55.00 &286.29 &41.97 &\underline{118.38} &\underline{23.33} &153.12 &-&-&\textbf{16.39} &\textbf{104.54}   \\
    & 49 &-&-&-&-&\underline{45.01} & \underline{126.91} &47.78 &276.42 &-&-&\textbf{17.50} &\textbf{111.54}  \\
    \hline

    \hline
    \multirow{2}{*}{0.5} 
    & 58 & 34.93 &64.80 & \underline{8.33} & \underline{43.16} &54.56 &173.30 &9.72 &60.97 &82.56 &404.45 &\textbf{7.50} & \textbf{38.67}  \\
    & 63 & -&-& \underline{24.44} & \underline{127.02} &-&-&40.83 &249.51 &-&-&\textbf{9.45} &\textbf{48.34}   \\
    \hline

    \hline
    \multirow{2}{*}{1.0} 
    & 64 & 39.01 &74.35 & \underline{11.67} & \textbf{61.03} &-&-&15.00 &92.64 &-&-&\textbf{10.56 }& \underline{66.62}   \\
    & 70 & -&-&\underline{45.56} & \underline{237.75} &-&-&-&-&-&-&\textbf{16.67} &\textbf{105.04}   \\
    \hline

    % \hline
    % \multirow{2}{*}{IID} 
    % & 40 & 0 & 0 & 0 & 0 & 0 & 0 & 0 & 0 & 0 & 0 & 0 & 0  \\
    % & 45 & 0 & 0 & 0 & 0 & 0 & 0 & 0 & 0 & 0 & 0 & 0 & 0  \\
    % \hline
    
    \end{tabular}%
}
% \vspace{-0.2in}
\end{table*}%

%% file: D1Table.tex
\begin{table}[htbp]
% \vspace{-0.1in}
  \caption{ABlation study for container selection and resource scheduling strategy}
  % \vspace{-0.1in}
  \label{D1Table}
  % \resizebox{\textwidth}{!}{
  \centering
    \begin{tabular}{|c|c|c|c|c|}
    \hline
     \multirow{3}{*}{Method} & \multicolumn{4}{c|}{Target Accuracy} \\
     \cline{2-5}
      & \multicolumn{2}{c|}{72\%} & \multicolumn{2}{c|}{73\%} \\
      \cline{2-5}
      &  Time (h) & Cost (\$) &  Time (h) & Cost (\$) \\
    \hline
     Ranodm & 36.11 & 176.89 &66.11 & 324.31 \\
    \hline
     NebulaFL w RR & 31.94 & 156.44 & 51.11 & 250.44 \\
    \hline
     NebulaFL w RC & 26.67 & 167.88 & 38.34 & 241.33 \\
    \hline
     NebulaFL wo C & 23.33 &157.94 & 30.56&206.8 \\
    \hline
     NebulaFL wo T & 33.62 &177.94 & 70.83 & 375.01 \\
    \hline
     NebulaFL & \textbf{23.33} & \textbf{146.91} & \textbf{29.45} & \textbf{185.28} \\
    \hline
    
    \end{tabular}%
% }
% \vspace{-0.15in}
\end{table}%

%% file: conclusion.tex
\section{Conclusion}

In this paper, we presented NebulaFL, a novel asynchronous Federated Learning framework designed to address the unique challenges of collaborative FL in JointCloud scenarios. NebulaFL effectively mitigates the issues of data heterogeneity, high communication overhead, and resource scheduling inefficiencies that typically hinder FL performance. Our approach incorporates a version control-based asynchronous training scheme to balance training times, a decentralized model rotation mechanism to facilitate efficient knowledge sharing among data centers and a reward-guided strategy for optimized container selection and resource scheduling. Extensive experimental evaluations demonstrate that NebulaFL not only achieves superior accuracy compared to existing FL methods but also significantly reduces communication costs and training time. In the future, our code will be further open-sourced and applied to real-world platforms, aiming to continually advance federated learning technologies and their practical applications.